\pdfoutput=1
\documentclass{aa}  
\usepackage{graphicx}
\usepackage{txfonts}
\usepackage{natbib} 
\usepackage{float}
\bibpunct{(}{)}{;}{a}{}{,} 
\begin{document}
   \title{Identification of transitional disks in Chamaeleon with \textit{Herschel}\thanks{{\it Herschel} is an ESA space observatory with science instruments provided by European-led Principal Investigator consortia and with important participation from NASA.}}

%
   \author{\'A. Ribas\inst{\ref{esac},\ref{cab},\ref{insa}}
     \and
     B.~Mer\'{i}n\inst{\ref{herschel}}
     \and
     H.~Bouy\inst{\ref{cab}} 
     \and
     C.~Alves~de~Oliveira\inst{\ref{esac}} 
     \and
     D.~R.~Ardila\inst{\ref{nasa}} 
     \and
     E.~Puga\inst{\ref{herschel}} 
     \and
     \'A.~K\'osp\'al\inst{\ref{estec}} 
     \and
     L.~Spezzi\inst{\ref{eso}}     
     \and
     N.~L.J.~Cox\inst{\ref{leuven}}
     \and
     T.~Prusti\inst{\ref{estec}} 
     \and
     G.~L.~Pilbratt\inst{\ref{estec}} 
     \and
     Ph.~Andr\'e\inst{\ref{saclay}} 
     \and
     L.~Matr\`a\inst{\ref{dublin}} 
     \and
     R.~Vavrek\inst{\ref{herschel}} 
}

   \offprints{\'A. Ribas}
	\institute{
	ESAC-ESA, P.O. Box, 78, 28691 Villanueva de la Ca\~{n}ada, Madrid, Spain\label{esac} \\
              \email{aribas@cab.inta-csic.es}
         \and
        Centro de Astrobiolog\'{i}a, INTA-CSIC, P.O. Box - Apdo. de correos 78, Villanueva de la Ca\~nada Madrid 28691, Spain\label{cab} 
        \and
	Ingenier\'ia y Servicios Aeroespaciales-ESAC, P.O. Box, 78, 28691 Villanueva de la Ca\~{n}ada, Madrid, Spain\label{insa}
        \and
	Herschel Science Centre, ESAC-ESA, P.O. Box, 78, 28691 Villanueva de la Ca\~{n}ada, Madrid, Spain\label{herschel}
        \and
        NASA Herschel Science Center, California Institute of Technology, 1200 E. California Blvd., Pasadena, CA 91125, USA\label{nasa}       
        \and
        Research and Scientific Support Department, ESTEC-ESA, PO Box 299, 2200 AG, Noordwijk, The Netherlands\label{estec} 
        \and
        European Southern Observatory, Karl-Schwarzschild-Strasse 2, 85748, Garching bei München, Germany\label{eso}
        \and
        Instituut voor Sterrenkunde, KU Leuven, Celestijnenlaan 200D, B-3001, Leuven, Belgium\label{leuven} 
        \and
        Laboratoire AIM Paris -- Saclay, CEA/DSM -- CNRS -- Universit\'e Paris Diderot, IRFU, Service d'Astrophysique, Centre d'Etudes de Saclay, Orme des Merisiers, 91191 Gif-sur-Yvette, France\label{saclay}
        \and
        School of Physics, Trinity College Dublin, Dublin 2, Ireland\label{dublin}
}

   \date{Received 19 December 2012; accepted 11 March 2013}

  \abstract
   {Transitional disks are circumstellar disks with inner holes that in some cases are produced by planets and/or substellar companions in these systems. For this reason, these disks are extremely important for the study of planetary system formation.}
   {The \textit{Herschel} Space Observatory provides an unique opportunity for studying the outer regions of protoplanetary disks. In this work we update previous knowledge on the transitional disks in the Chamaeleon I and II regions with data from the \textit{Herschel} Gould Belt Survey.} 
   {We propose a new method for transitional disk classification based on the WISE 12\,$\mu$m $-$ PACS 70\,$\mu$m color, together with inspection of the \textit{Herschel} images. We applied this method to the population of Class II sources in the Chamaeleon region and studied the spectral energy distributions of the transitional disks in the sample. We also built the median spectral energy distribution of Class II objects in these regions for comparison with transitional disks.}
   {The proposed method allows a clear separation of the known transitional disks from the Class II sources. We find six transitional disks, all previously known, and identify five objects previously thought to be transitional as possibly non-transitional. We find higher fluxes at the PACS wavelengths in the sample of transitional disks than those of Class II objects.}
   {We show the \textit{Herschel} 70\,$\mu$m band to be a robust and efficient tool for transitional disk identification. The sensitivity and spatial resolution of \textit{Herschel} reveals a significant contamination level among the previously identified transitional disk candidates for the two regions, which calls for a revision of previous samples of transitional disks in other regions. The systematic excess found at the PACS bands could be either a result of the mechanism that produces the transitional phase, or an indication of different evolutionary paths for transitional disks and Class II sources.}


   \maketitle


\section{Introduction}

Protoplanetary disks surrounding young stars are known to evolve over timescales of a few million years from a more massive and optically thick phase (Class II objects) to optically thin debris disk systems \citep[Class III sources; see ][for a recent review on the evolution of protoplanetary disks]{Williams2011}. There are several indications of this evolution with time. Infrared (IR) observations of star-forming regions show a systematic decrease of the IR flux with stellar age \citep{Haisch2001,Gutermuth2004,Sicilia-Aguilar2006,Currie2009}. In the optical and ultraviolet, observations show that the disk mass accretion rate decreases with time as predicted by disk evolutionary models \citep{Hartmann1998,Calvet2005,Fedele2010,Sicilia-Aguilar2010,Spezzi2012}. Another important evidence is found in deep mid-IR spectroscopic observations of young stars with disks that show dust grain growth, crystallization, and settling to the disk mid-plane. These phenomena are found to be correlated with the evolution of the disk structure across two orders of magnitude in stellar mass \citep{Meeus2001,vanBoekel2005,Kessler-Silacci2005,Apai2005,Olofsson2009}.

Most of the evolution of protoplanetary disks is driven by gravitational interaction and viscosity effects in the disk \citep{Pringle1981}. However, some circumstellar disks show evidence of a different evolutionary phase: they are known as transitional disks. Compared to Class II disks, they display a clear dip in their spectral energy distribution (SED) at short-mid IR (typically around 8-12 $\mu$m) and a rising SED with flux excesses similar to that of Class II sources at longer wavelengths \citep{Strom1989,Calvet2002,Espaillat2007b,Andrews2011}. The dips in the SEDs are usually explained in terms of dust-depleted regions and/or cavities in the disks, of typical sizes of some tens of AU \citep[see][and references therein]{Merin2010,Andrews2011}.

Several processes have been proposed to explain these gaps and holes: gravitational interaction with a low-mass companion \citep{Bryden1999,Rice2003a,Papaloizou2007}, photo-evaporation \citep{Clarke2001,Alexander2006a,Alexander2006b}, or grain growth \citep{Dullemond2005,Tanaka2005,Birnstiel2012}. Observational evidence of stellar or substellar companions has been obtained in some cases \citep[i.e., CoKu Tau4 or T Cha, see][respectively]{Ireland2008,Huelamo2011}. If we were able to distinguish between these different explanations would better understand the mechanisms that produce the gaps in transitional disks, and the planetary formation scenario. For this reason, any hint on which process governs the transition phase is relevant.

In this paper, we investigate the contribution of the far-IR data from the \textit{Herschel} Space Observatory \citep{Herschel} to our understanding of transitional disks. We present a new method for transitional disk identification and apply it to the sample of Class II objects in the Chamaeleon (Cha) I and II regions. Section~\ref{observations} describes the data reduction process, the sample selection, and the photometry extraction. Section~\ref{identifying} explains the proposed method used in the paper to identify and reclassify transitional disks. A more detailed discussion of the sample of transitional disks is given in Sect.~\ref{outerdisks}. Section~\ref{conclusions} summarizes our results.

\section{Observations and sample}\label{observations}

\subsection{Observations}

The Cha I and II regions were observed by the \textit{Herschel} Space Observatory in the context of the Gould Belt Survey \citep{Andre2010}. These regions are part of the Chamaeleon molecular cloud complex that also includes the Cha III cloud. The complex is located at 150-180\,pc \citep{Whittet1997} and is one of the most often studied low-mass star-forming regions because of its proximity. Cha I has an estimated age of $\sim$2 Myr and a population of $\sim$200 young stellar objects \citep{Luhman2008a,Winston2012}. Cha II harbors a smaller population ($\sim$ 60) of young sources \citep{Young2005,Spezzi2008}. Because of their age and location, these regions are perfect scenarios for transitional disk search and study.

Two sets of observations were used for each region: a first set taken in parallel mode, using the PACS \citep[70 and 160,\,$\mu$m][]{PACS} and SPIRE \citep[250\, 350, and 500\,$\mu$m,][]{SPIRE} instruments at a speed of 60\arcsec/s, and the 100\,$\mu$m PACS band at 20\arcsec/s from a second set in scan mode. The observing strategy is described in more detail in \citet{Andre2010}. The total observing time in parallel mode for Cha I was $\sim$\,8 hours and 6 hours for Cha II, covering a total area of $\sim$\,9\,deg$^2$ ($\sim$\,5.5 and 3.5\,deg$^2$). The PACS 100\,$\mu$m images covered 2.6\,deg$^2$ in Cha I and 2\,deg$^2$ in Cha II, and add up to a total time of 8 hours and 6 hours, respectively \citep[see][
and Spezzi et al.\,in prep. for a detailed description of the data sets]{Winston2012}. Obsids for Cha I are 1342213178, 1342213179 (parallel mode) and 1342224782,1342224783 (scan mode), and obsids 1342213180,1342213181 (parallel mode), and 1342212708, 1342212709 (scan mode) for Cha II.

\begin{figure*}
  \includegraphics[width=.97\textwidth]{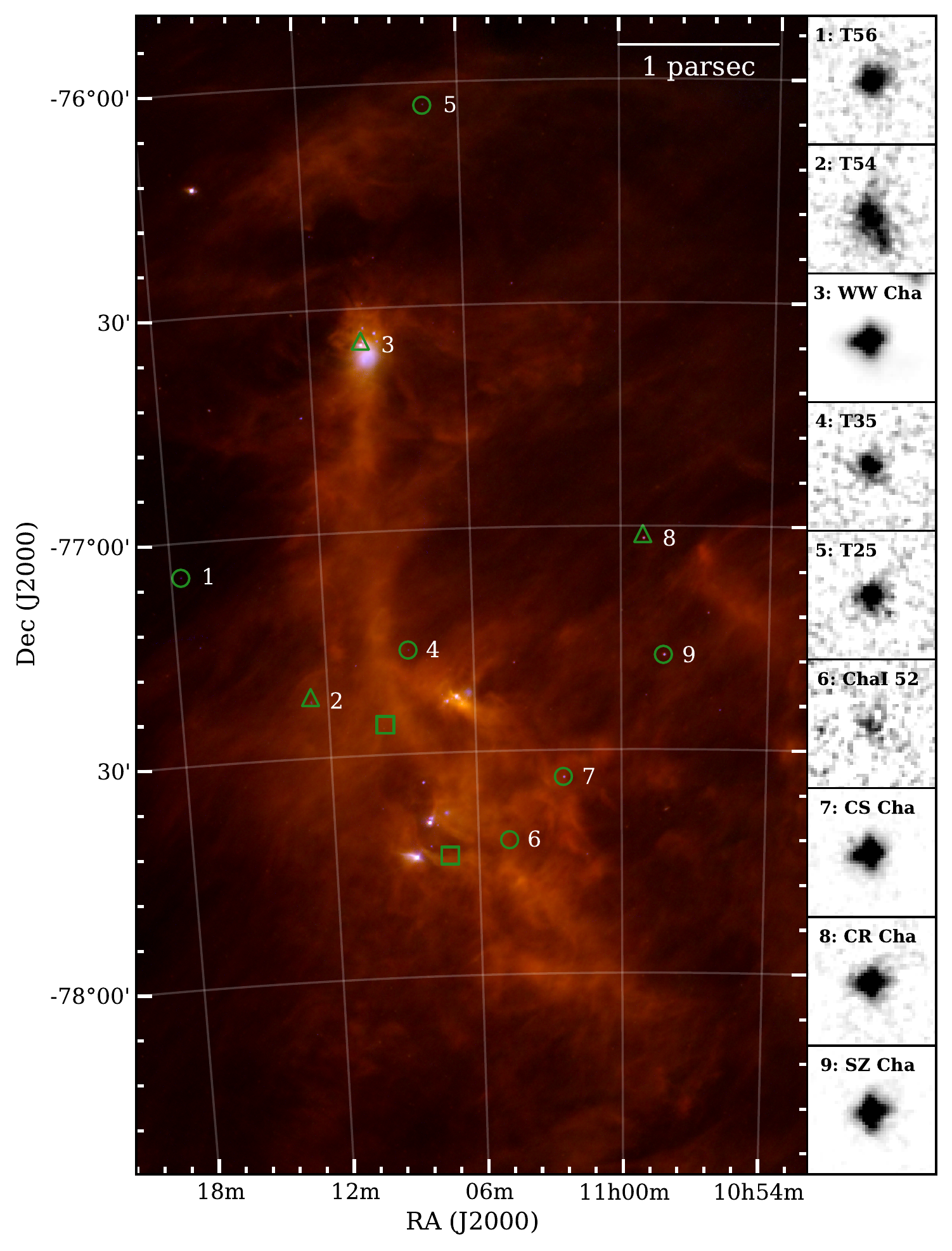}
  \caption{Left: Three-color composite image of the Cha I region (blue: PACS\,70\,$\mu$m, green: PACS\,160\,$\mu$m, red: SPIRE\,250\,$\mu$m).  Circles mark the position of transitional disks detected in the \textit{Herschel} images and classified with the proposed method (see Sec.\ref{sec:classification}). Triangles show sources not fulfilling our selection criteria, and squares represent non-detected sources. Right: thumbnails of the 70\,$\mu$m \textit{Herschel} maps (50\,\arcsec\,$\times$\,50\,\arcsec). The color scale ranges from the median value (background level) to 5\,$\sigma$ over this value (black). In both figures, north is up, and east is left. Note that WW Cha is located on a bright core.}\label{fig:cha}
\end{figure*}

The data were pre-processed using the \textit{Herschel} interactive processing environment \citep[HIPE,][]{Ott2010} version 9. The final maps were created using \textit{Scanamorphos} \citep{Roussel2012} for PACS, and the \textit{destriper} algorithm included in HIPE for  SPIRE. These two algorithms are optimized for regions such as Chamaeleon, which have bright extended emission.
Figure~\ref{fig:cha} shows a three-color composite image of Cha I (blue: 70\,$\mu$m, green: 160\,$\mu$m, and red: 250\,$\mu$m).

\subsection{Sample selection}

\citet{Luhman2008a} and \citet{Luhman2008b} presented the largest census of young stellar objects (YSOs) members of Cha I including \textit{Spitzer} photometry, and \citet{Alcala2008} and \citet{Spezzi2008} did the same for Cha II. We selected from these studies the sources classified as Class II with known extinction values. Since we aim to classify transitional disks, we also included T25, flagged as Class III in \citet{Luhman2008a} but later found to be a transitional source in \citet{Kim2009}. We rejected objects with signal-to-noise ratio (S/N) $<$\, 5 in any of the 2MASS bands to ensure a good photometry estimation and coordinates measurement. The final sample of Class II objects is comprised of 119 sources.

To our knowledge, 12 sources in the sample are classified as transitional disk candidates in the literature: SZ Cha, CS Cha, T25, T35, T54, T56, and CHXR 22E from \citet{Kim2009}, C7-1 from \citet{Damjanov2007}, CR Cha, WW Cha, and ISO-ChaI 52 from \citet{Espaillat2011}, and ISO-ChaII 29 from \citet{Alcala2008}.

\subsection{Photometry}

We extracted \textit{Herschel} photometry of the Class II sample following these steps:

\begin{enumerate}

\item We used the \textit{Sussextractor}  algorithm \citep{Savage2008} in HIPE to detect sources with an S/N $>$ 5 in the PACS images. We then visually checked that no obvious source was missing.

\item We cross-matched the initial sample with the detections in the PACS images, using a search radius of 5\arcsec. This radius was chosen based on the size of the point spread functions (PSFs) at these wavelengths ($\sim$5.8\arcsec $\times$ 12.1\arcsec, 6.7\arcsec $\times$ 7.3\arcsec, and 11.4\arcsec $\times$ 13.4\arcsec, for the 70, 100, and 160\,$\mu$m bands at the corresponding observing speeds). We note that the background emission becomes more sifnificant for longer wavelengths, producing false detections because of bright filaments and ridges. To avoid possible mismatches, we considered as \textit{Herschel}-detected sources only those with counterparts at any PACS band. For SPIRE, we found the \textit{Sussextractor} output to be highly contaminated with false detections. Therefore, we  visually inspected the positions of the detected sources individually for these bands.

\item We performed aperture photometry centered on the 2MASS coordinates of each detection. We used the recommended aperture radii and background estimation annulus for each band (see the PACS point-source flux calibration technical note from April 2011, and Sect. 5.7.1.2 of the SPIRE data reduction guide). The values for the apertures, inner and outer annulus radii (in this order) are 12\arcsec, 20\arcsec, 30\arcsec for 70 and 100\,$\mu$m, 22\arcsec, 30\arcsec, 40\arcsec for 160\,$\mu$m, 22\arcsec, 60\arcsec, 90\arcsec for 250\,$\mu$m, 30\arcsec, 60\arcsec, 90\arcsec for 350\,$\mu$m, and 42\arcsec, 60\arcsec, 90\arcsec for 500\,$\mu$m.

\item Since aperture photometry was used, objects close to bright filaments or cores are likely to suffer from contamination. Also, given the size of the PSF, no photometric measurements can be performed for close objects (separation less than $\sim$\,10\arcsec). Therefore we rejected ten detections that showed obvious problems in their photometry or enclosed more than one object.

\end{enumerate}

After excluding the transitional disks, the final result of this process is 41 Class II sources detected at any PACS band  (26 and 15 in Cha I and Cha II, respectively), nine of them detected also with SPIRE.

We checked that the obtained photometry was consistent with that from other map-making algorithms (such as \textit{photProject} for PACS), and found no significant deviation.

We visually inspected the position of non-detected transitional disks in the original sample and found that C7-1, CHXR 22E, and ISO-ChaII 29 are not detected at any of the \textit{Herschel} bands. Additionally, ISO-ChaI 52 is not detected by \textit{Sussextractor} at 70\,$\mu$m, but it is at 100\,$\mu$m. The object is visually found in the 70\,$\mu$m image with a flux of 150\,mJy over a background root mean square (RMS) of 40\,mJy. We therefore included the 70\,$\mu$m flux in our analysis. Source T25 is not in the field of view of the 100\,$\mu$m map, which is smaller than the parallel mode observations. Coordinates and stellar parameters for the transitional disks in this study can be found in Table~\ref{table:params}.

\subsection{Photometric uncertainties and upper limits}

The absolute calibration errors for PACS and SPIRE are 5\,\%
 and 7\,\%
, respectively (see PACS and SPIRE observer manuals). To ensure a conservative error estimation, we used a 15\,\% error value for PACS and 20\,\% for SPIRE, taking into account that the background emission becomes increasingly stronger at longer wavelengths. 

When no source was detected, we computed an upper limit calculating the RMS of 100 apertures taken around the source, using the same aperture radii and correction factors as for the detections. The extracted PACS and SPIRE fluxes for the 12 transitional candidates in the considered sample are reported in Table~\ref{table:herschel}.

\begin{table*}
  \caption{Coordinates and stellar parameters of the 12 transitional disk candidates analyzed in this work.}\label{table:params}
  \begin{center}
    \begin{tabular}{l c c c c c c c c c c c c c c c}
      \hline
      \hline\rule{0mm}{3mm}Name & R.A.$_{J2000}$ & Dec.$_{J2000}$ & A$_V$ & SpT & T$_*$ & L$_*$ & M$_*$ & R$_*$ & Refs.\\
      & & & (mag) & Type & (K) & (L$_\odot$) & (M$_\odot$) & (R$_\odot$)
      \vspace{1.mm}
      \\\hline \object{CR Cha} & 10:59:06.97 & -77:01:40.3 & 1.5& K2 & 4900 & 3.5 & 1.9 & 2.6 & 1,2,3,4,5,6 \\
      \object{CS Cha} & 11:02:24.91 & -77:33:35.7 & 0.25 & K6 & 4205 & 1.5 & 0.9 & 2.3 & 1,2,3,5,6,7,8,9 \\
      \object{SZ Cha} & 10:58:16.77 & -77:17:17.1 & 1.90 & K0 & 5250 & 1.9 & 1.4 & 1.7 & 1,2,3,4,5,6,7,8,9 \\
      \object{WW Cha} & 11:10:00.11 & -76:34:57.9 & 4.8 & K5 & 4350 & 6.5 & 1.2 & 4.5 & 1,2,3,4,5,6,7,8 \\
      \object{T25} & 11:07:19.15 & -76:03:04.9 & 0.78 & M3 & 3470 & 0.3 & 0.3 & 1.5 & 1,2,3,4,5,6,7,9 \\
      \object{T35} & 11:08:39.05 & -77:16:04.2 & 3.5 & M0 & 3850 & 0.4 & 0.6 & 0.5 & 1,2,3,4,5,6,7,9 \\
      \object{T54} & 11:12:42.69 & -77:22:23.1 & 1.78 & G8 & 5520 & 4.1 & 2.4 & 1.5 & 1,2,3,4,5,6,7,9,10,11 \\
      \object{T56} & 11:17:37.01 & -77:04:38.1 & 0.23 & M0.5 & 3720 & 0.4 & 0.5 & 1.6 & 1,2,3,4,5,6,7,9 \\
      \object{ISO-ChaI 52} & 11:04:42.58 & -77:41:57.1 & 1.3 & M4 & 3370 & 0.1 & 0.3 & 1.0 & 2,3,4,5,7 \\
      \object{C7-1} & 11:09:42.60 & -77:25:57.9 & 5.0 & M5 & 3125 & \ldots & \ldots & \ldots & 3,4,5,7,12 \\
      \object{CHXR 22E} & 11:07:13.30 &-77:43:49.9 & 4.79 & M3.5 & 3400 & 0.2 & \ldots & \ldots & 3,5,7,9 \\
      \object{ISO-ChaII 29} & 12:59:10.14 & -77:12:13.9 & 5.57 & M0 & 3850 & 0.65 & \ldots & 1.85 & 13,14
      \\\hline

    \end{tabular}
    \tablebib{(1)~\citet{Gauvin1992}; (2)~\citet{Espaillat2011};  (3)~\citet{Luhman2007}; (4)~\citet{Luhman2008b}; (5)~\citet{Manoj2011}; (6)~\citet{Henning1993}; (7)~\citet{Luhman2008a}; (8)~\citet{Belloche2011}; (9)~\citet{Kim2009}; (10)~\citet{Lafreniere2008}; (11)~\citet{Preibisch1997}; (12)~\citet{Damjanov2007}; (13)~\citet{Spezzi2008}; (14) ~\citet{Alcala2008}}

  \end{center}
\end{table*}

\begin{table*}
  \begin{center}
   \caption{\textit{Herschel} photometry of the 12 transitional disks in the sample.}\label{table:herschel}
    \begin{tabular}{l c c c c c c }
      \hline
      \hline\rule{0mm}{3mm}Name & F$_{\rm 70 \mu m}$ & F$_{\rm 100 \mu m}$ & F$_{\rm 160 \mu m}$ & F$_{\rm 250 \mu m}$ & F$_{\rm 350 \mu m}$ & F$_{\rm 500 \mu m}$ \\
      & (Jy) & (Jy) & (Jy) & (Jy) & (Jy) & (Jy)
      \vspace{1.mm}
      \\\hline
      \multicolumn{7}{c}{Detected sources}
      \\\hline \object{CR Cha} & 1.61\,$\pm$\,0.24 & 2.19\,$\pm$\,0.33 & 2.74\,$\pm$\,0.41 & 2.37\,$\pm$\,0.47 & 1.69\,$\pm$\,0.34 & 1.09\,$\pm$\,0.22\\
      \object{CS Cha*} & 3.08\,$\pm$\,0.46 & 2.82\,$\pm$\,0.42 & 2.32\,$\pm$\,0.35 & 0.88\,$\pm$\,0.18 & 0.38\,$\pm$\,0.08 & 0.13\,$\pm$\,0.03 \\
      \object{SZ Cha} & 3.88\,$\pm$\,0.58 & 3.63\,$\pm$\,0.54 & 3.86\,$\pm$\,0.58 & 2.85\,$\pm$\,0.57 & 1.94\,$\pm$\,0.39 & 1.14\,$\pm$\,0.23 \\
      \object{WW Cha} & 25.91\,$\pm$\,3.88 & 32.32\,$\pm$\,4.85 & 27.3\,$\pm$\,4.10 & 24.92\,$\pm$\,4.99 & 12.44\,$\pm$\,2.49 & 6.79\,$\pm$\,1.36 \\
      \object{T25} & 0.52\,$\pm$\,0.08 & \ldots & 0.50\,$\pm$\,0.08 & 0.20\,$\pm$\,0.04 & 0.11\,$\pm$\,0.02 & $<$\,0.10 \\
      \object{T35} & 0.38\,$\pm$\,0.06 & 0.36\,$\pm$\,0.06 & 0.200\,$\pm$\,0.03 & $<$\,1.69 & $<$\,2.10 & $<$\,2.06 \\
      \object{T54} & 0.60\,$\pm$\,0.09 & 0.77\,$\pm$\,0.12 & 0.98\,$\pm$\,0.15 & 0.46\,$\pm$\,0.09 & $<$\,1.04 & $<$\,1.18 \\
      \object{T56} & 0.68\,$\pm$\,0.10 & 0.57\,$\pm$\,0.09 & 0.30\,$\pm$\,0.05 & 0.30\,$\pm$\,0.05 & 0.30\,$\pm$\,0.06 & 0.11\,$\pm$\,0.02 \\
      \object{ISO-ChaI 52} & 0.15\,$\pm$\,0.02 & 0.15\,$\pm$\,0.02 & $<$\,1.07 & $<$\,1.42 & $<$\,2.06 & $<$\,2.04
      \\\hline
      \multicolumn{7}{c}{Undetected sources}
      \\\hline
      \object{C7-1} & $<$\,0.04 & $<$\,0.08 & $<$\,0.94 & $<$\,1.24 & $<$\,1.69 & $<$\,2.10\\
      \object{CHXR 22E} & $<$\,0.08 & $<$\,0.14 & $<$\,1.10 & $<$\,1.19 & $<$\,1.18 & $<$\,0.96 \\
      \object{ISO-ChaII 29} & $<$\,0.04 & $<$\,0.07 & $<$\,0.85 & $<$\,1.41 & $<$\,2.65 & $<$\,3.00 \\\hline
    \end{tabular}
   \end{center}
  * SPIRE photometry is very likely contaminated for this source (see appendix).
\end{table*}

\section{Identification of transitional disks}\label{identifying}

\subsection{Photometric selection}\label{photselection}

Several selection criteria have been used in the past to separate transitional disks from Class II sources. \citet{Fang2009} used a color-color diagram based on the \textit{Ks} band and on the [5.8], [8.0] and [24] \textit{Spitzer} bands. \citet{Muzerolle2010} proposed a classification criterion based on the slope of the SED between 3.6 and 4.8\,$\mu$m and between 8 to 24\,$\mu$m. \citet{Cieza2010} also used a color-color diagram, based on the \textit{Spitzer} photometry at 3.6, 4.5 and 24 $\mu$m. However, all these methods were found to suffer from different contamination levels, as explained in \citet{Merin2010}.

There is a high diversity in the morphology of transitional disks, hence there are various definitions. However, most of them share two common characteristics: (1) they have low or no excess with respect to the photosphere up to the $\lambda_{\rm turn-off}$ or the pivot point, usually found around $\sim$ 8-10\,$\mu$m, and (2) they have strong excesses for longer wavelengths \citep[see section 7.1 in][and references therein]{Williams2011}. This is translated into a decreasing slope of the SED up to $\lambda_{\rm turn-off}$, and an increasing one for longer wavelengths. 

To identify transitional disks using \textit{Herschel} photometry, we computed two spectral indexes ($\alpha$): one between the \textit{Ks} band and 12\,$\mu$m ($\alpha_{Ks-12}$), and the other between 12\,$\mu$m and 70\,$\mu$m ($\alpha_{12-70}$). The spectral index is defined as $\alpha_{\lambda_1 - \lambda_2}=\frac{{\rm log}(\lambda_1 {\rm F}_{\lambda_1})-{\rm log}(\lambda_2 {\rm F}_{\lambda_2})} {{\rm log}(\lambda_1)-{\rm log}(\lambda_2)}$, where $\lambda$ is measured in $\mu$m and F$_{\lambda}$ in erg s$^{-1}$ cm$^{-2}$ s$^{-1}$. Therefore, $\alpha$ traces the slope of the SED in the considered range ($\alpha > 0 \rightarrow$ rising SED, $\alpha < 0 \rightarrow$ decreasing SED). This spectral index has been intensively used since its introduction by \citet{Lada1984} to classify protostars and young objects. 

\begin{figure}
  \centering
  \includegraphics[width=\hsize]{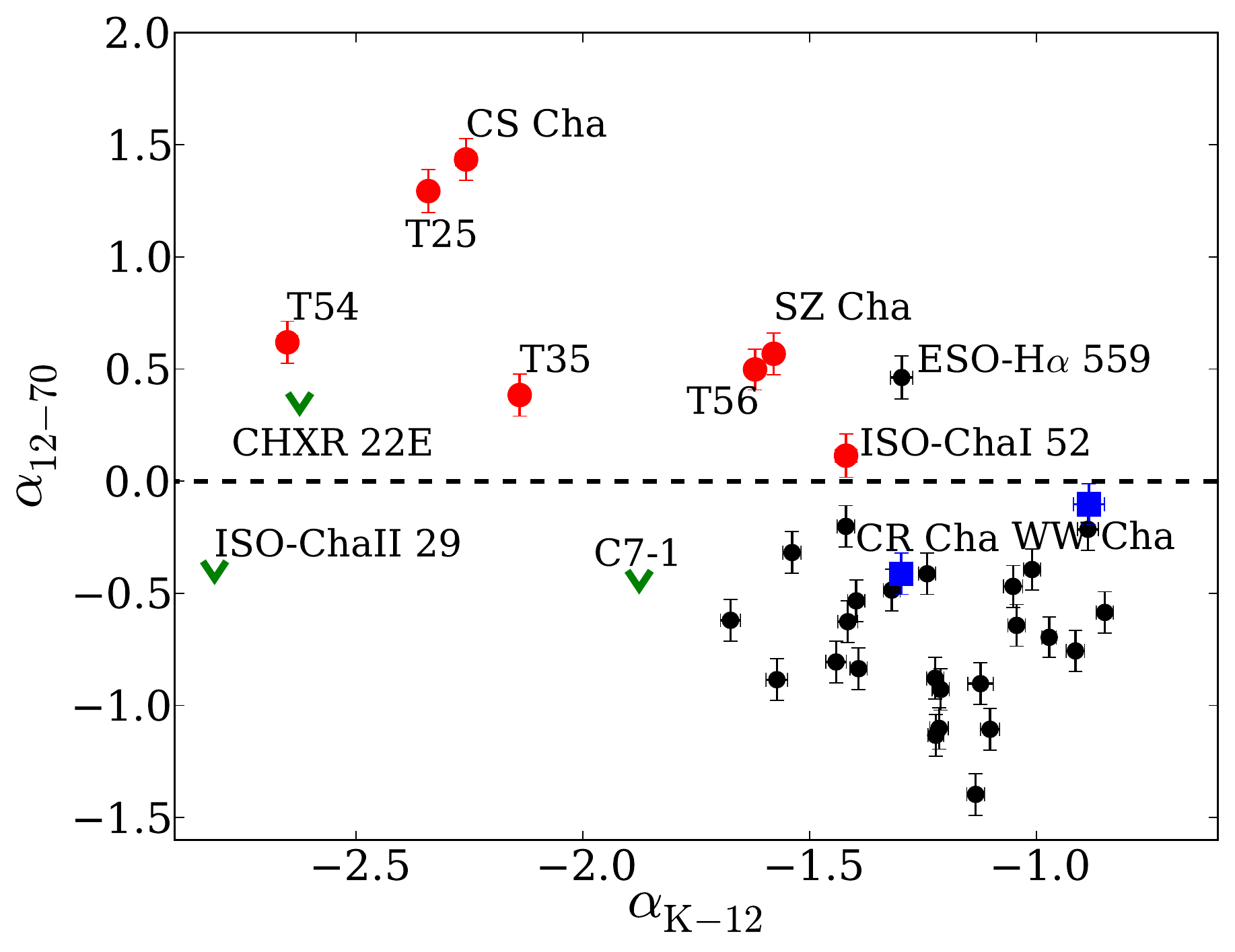}
  \caption{SED slope between the 12 and 70\,$\mu$m ($\alpha_{12-70}$) as a function of the SED slope between the \textit{Ks}-band and 12\,$\mu$m ($\alpha_{Ks-12}$). Transitional disks from the literature meeting the selection criterion are marked as red dots, green downward arrows are those for which only upper limits could be estimated. Class II objects are black dots. Blue squares are pretransitional disks from \citet{Espaillat2011}. There is a clear separation between Class II and transitional disks due to the different shape of their SED. The single black dot with $\alpha > 0 $ is ESO-H$\alpha$ 559, an edge-on disk. This diagram shows the potential for transitional disk classification using the 70\,$\mu$m band.}\label{fig:slopes}
\end{figure}

Figure~\ref{fig:slopes} demonstrates that these two slopes together efficiently separate the two populations. The separation is clearer in the 12-70\,$\mu$m axis, where $\alpha_{12-70} < 0$ corresponds to typical Class II sources, and $\alpha_{12-70} > 0$ is indicative of the transitional nature of the objects. This separation in the slope between 12\,$\mu$m and 70 $\mu$m is an expected feature: for short-mid IR wavelengths, the slope depends strongly on the presence of weak excess, or on the spectral type of the star if there is no excess. On the other hand, the definition of transitional disks itself guarantees a positive slope for longer wavelengths. This separation also reveals the usefulness of the \textit{Herschel} data for this classification. As a result of the selection method, two disks reported in \citet{Espaillat2011}, WW Cha and CR Cha, are not separated from Class II objects and we confirm that they do not deviate significantly from the median SED of the Class II sources in Cha I and II (Fig.~\ref{fig:sedsmiss}). Based on this evidence, we consider them as non-transitional. The rest of the transitional disks are properly separated from Class II sources. The computed upper limits also allow us to classify C7-1 and ISO-ChaII 29 as non-transitional using this method.

Interestingly, one Class II source shows $\alpha_{12-70} > 0$ in the former diagram. The object, called ESO-H$\alpha$ 559, has been recently identified as a probable edge-on disk in \citet{Robberto2012} by modeling its SED. Its underluminosity with respect to its spectral type also supports this scenario. We find this type of object to be a source of contamination for this method: edge-on disks can mimic the SED of transitional sources. Their geometry will cause a high circumstellar extinction level, blocking the light from the central star at short wavelengths \citep{Stapelfeldt1999,Wood2002,Duchene2010,Huelamo2010}. The disk becomes optically thin for longer wavelengths ($>$\,24\,$\mu$m) and the emission from the star can pass through the disk, resulting in an increase of the flux and hence a positive slope of the SED. When their spectral type is known, edge-on disks can be identified in Hertzsprung–Russell diagrams, as they are often underluminous.

We also note that this method is not suitable for detecting a small subsample of transitional disks called anemic \citep{Lada2006}, homologously depleted \citep{Currie2009}, or weak excess disks \citep{Muzerolle2010}. They are defined as objects with low excess at all infrared wavelengths and show $\alpha_{\rm excess} < 0$. For this reason, they cannot be found with the criterion proposed in this work. On the other hand, the rest of transitional disks should display $\alpha_{12-70} > 0$ and hence can be properly separated.

\subsection{Morphological classification}

We checked whether any of the remaining seven transitional disks were spatially resolved in the \textit{Herschel} images. Extended emission could indeed indicate contamination by a coincident background source, a close by object, or the extended background emission, as shown in \citet{Matra2012}. 

Given the estimated distance of 160\,pc to the Cha I molecular cloud \citep{Whittet1997,Luhman2008b}, the full-width at half-maximum of the PSFs for the three PACS bands ($\sim$ 6\arcsec, 7\arcsec and 12\,\arcsec, see the official PACS PSF document\footnote{http://herschel.esac.esa.int/twiki/pub/Public/PacsCalibrationWeb/ bolopsf\_20.pdf}) would allow us to resolve structures of $\sim$\,900-2000\,AU. It is difficult to define an outer radius for protoplanetary/transitional disks, but typical values range from some tens to $\sim$\,1000\,AU in the most extreme cases \citep{Williams2011}. Direct imaging of proplyds and disks in the Trapezium cluster by \citet{Vicente2005} showed the size distribution to be contained within 50 and 100\,AU. On the other hand, the R$_{c}$ parameter \citep[defined as the radius where the surface density deviates significantly from a power law and the disk density declines rapidly, see][]{Williams2011} has typical values between 15-230\,AU \citep{Hughes2008,Andrews2009,Andrews2010b}. 

\begin{figure}
  \centering
  \includegraphics[width=\hsize]{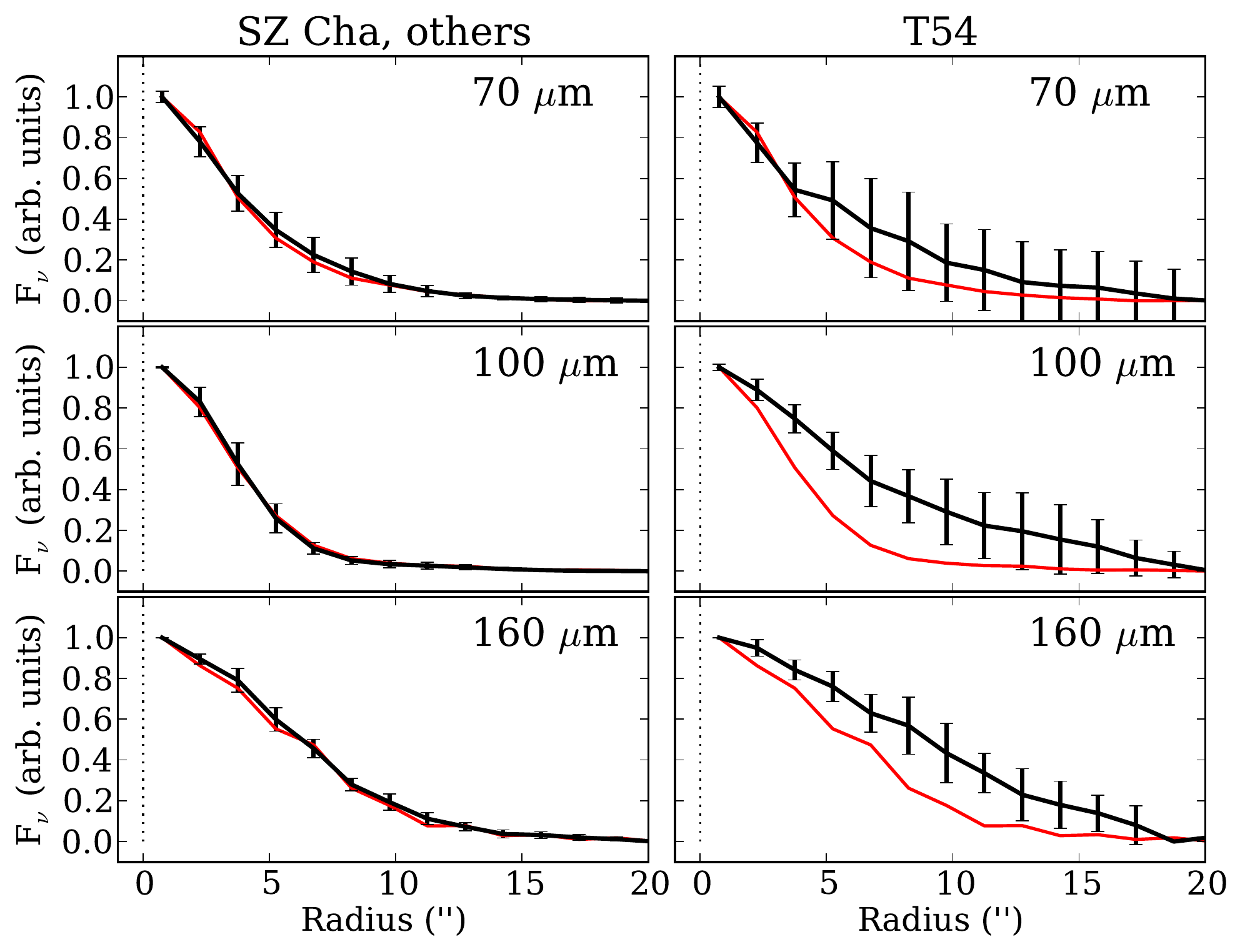}
  \caption{Average brightness radial profile (black line) for a point source (SZ Cha, left) and an extended source (T54, right) compared to the observational PSF radial profile (red line). The error bars are the RMS of the values. All considered sources present the same behavior as SZ Cha except for T54, whose observed radial profile is clearly above the PSF profile, indicating that this source is extended and all others are point-like in \textit{Herschel} images.}\label{fig:radial}
\end{figure}

This suggests that none of these sources should be resolved in the \textit{Herschel} images. In each of the PACS band, we compared the radial profile of the transitional disks with an empirical PSF constructed using clean isolated point sources. Of the nine detected sources, only T54 was found to be extended, as shown in Fig.~\ref{fig:radial}. \citet{Matra2012} showed that the excess beyond 100\,$\mu$m is likely not related to the source, and therefore not originating from a circumstellar disk. This interpretation results in a substantial decrease in the IR excess coming from T54, making its SED not representative of the characteristic inner-hole geometry around transitional disks. The case of T54 shows that one needs to verify the origin of the IR excess in protoplanetary/transitional disks.

We found no other extended sources in the \textit{Herschel} images and conclude that all the detected transitional disks have far-IR excesses related to the sources. Therefore all but one (T54) of the transitional disk candidates in the Cha I and II regions are confirmed to be point-sources, up to the resolution of the \textit{Herschel} PACS and SPIRE instruments.

\subsection{Transitional disk classification}\label{sec:classification}

Thanks to the new \textit{Herschel} photometric data, we are able to reclassify the already known transitional disks in the Cha I and II regions based on the shape of their SEDs. CS Cha, SZ Cha, T25, T35, T56, and ISO-ChaI 52 show a typical transitional disk SED.

Two objects from \citet{Espaillat2011} do not fulfill our selection criterion, which is tuned to identify clear signatures of inner holes. These sources were selected by \citet{Espaillat2011} based on their silicate feature strengths, and therefore the different results obtained in this study are not surprising. CR Cha shows weak excess up to 2\,$\mu$m and a typical Class II SED for longer wavelengths. The SED of WW Cha does not display any decrease in its IR emission, typical for transitional disks. Indeed, the \textit{Herschel} images support one of the scenarios proposed by \citet{Espaillat2011}: WW Cha is located on one of the cores in Cha I and presumably accretes at a high rate. Furthermore, it shows a strong excess along the whole wavelength range and therefore cannot be considered as a transitional disk, but ir more likely a Class II source. The dust-rich environment around WW Cha might also contaminate the \textit{Herschel} photometry and account for part of the observed IR excess emission. These sources are then probably non-transitional. Moreover, the morphological analysis of the candidates shows that T54 is extended (Fig.~\ref{fig:radial}) and hence a misclassified object. 

Conclusions for the non-detected sources are more complicated to draw, and they should be treated with caution, since a non-detection does not directly reject a candidate, but could simply be due to a sensitivity bias. The computed upper limits for C7-1 and ISO-ChaII 29 exclude them as transitional disks according our selection criterion. ISO-ChaII 29 is a special case: the upper limit of 35\,mJy in PACS 70\,$\mu$m is lower than the detection of 56.90$\pm$8.63\,mJy in the MIPS 70\,$\mu$m band indicated in \citet{Alcala2008}, and these two measurements are inconsistent. ISO-ChaII 29 shows both strong Li absorption and H$\alpha$ emission, which confirms it as a YSO \citep{Spezzi2008}. However, it is the only transitional disk in the sample with photospheric fluxes up to 24\,$\mu$m in our and the \citet{Alcala2008} sample. These authors also found it to have the steepest $\alpha_{excess}$. This object is therefore probably misclassified in the \textit{Spitzer} images, as strongly suggested by the non-detection in any of the \textit{Herschel} bands and by the outlier nature of the object if the MIPS detection is considered. We therefore reclassify it as a non-transitional source. We stress that our method is unable to detect transitional disks with weak excesses, and deeper observations should be made to confirm or reject the presence of disks and holes in these systems. In the case of CHXR 22E it is not possible to extend the analysis further without making strong assumptions about the disk mass and morphology. For this reason, we exclude it for the remainder of the study. 

As mentioned in Sect. \ref{photselection}, it is important to note that the proposed criterion will only select transitional objects with increasing slopes between 12 and 70\,$\mu$m. This feature is unlikely to be produced by grain growth alone \citep[see][for a review on the topic]{Williams2011}. As a result, the proposed classification criterion may be biased toward detecting only transitional disks with large inner holes produced by photoevaporation, gap opening by unresolved companions, giant planet formation, or a combination of these scenarios. Physical interpretation of this peculiar SED shapes requires detailed modeling, and there is no full consensus yet on which physical phenomena can be safely attributed to each type of SED \citep[see e.g.][]{Birnstiel2012}. A more detailed analysis of this topic will help to determine the real impact of this selection effect.

From the initial sample of 12 transitional disk candidates in the Cha I and II regions we confirm six objects to be transitional disks, reject five sources by photometric or morphological criteria, and leave one object unclassified since it is not detected in the \textit{Herschel} images. These numbers imply a significant ($\sim$\,45\,\%) observed contamination level in the transitional disk sample considered in this study. Given the small number statistics, it is premature to extend this result to other samples. However, this result calls for a revision of the known transitional disks: if applicable to the whole sample, this contamination level would imply a shorter transitional-phase lifetime and hence could shed some light on the mechanisms responsible for the evolution of protoplanetary disks.

\begin{figure*}
  \centering
  \includegraphics[width=0.32\hsize]{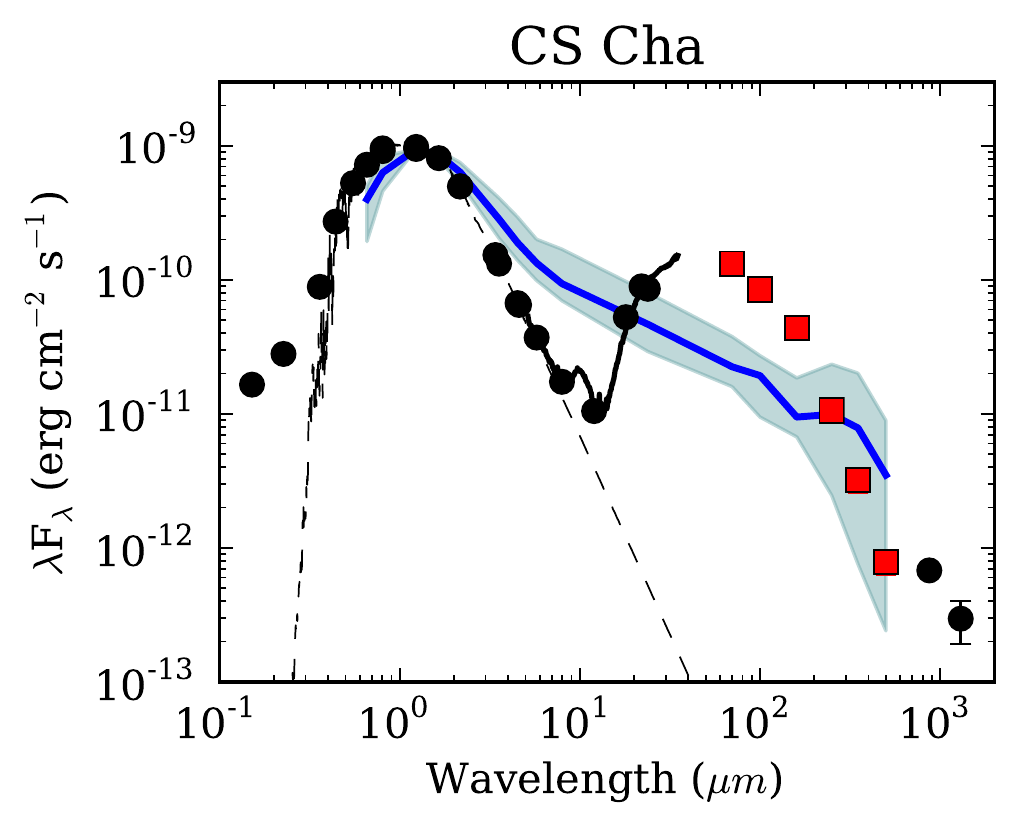}
  \includegraphics[width=0.32\hsize]{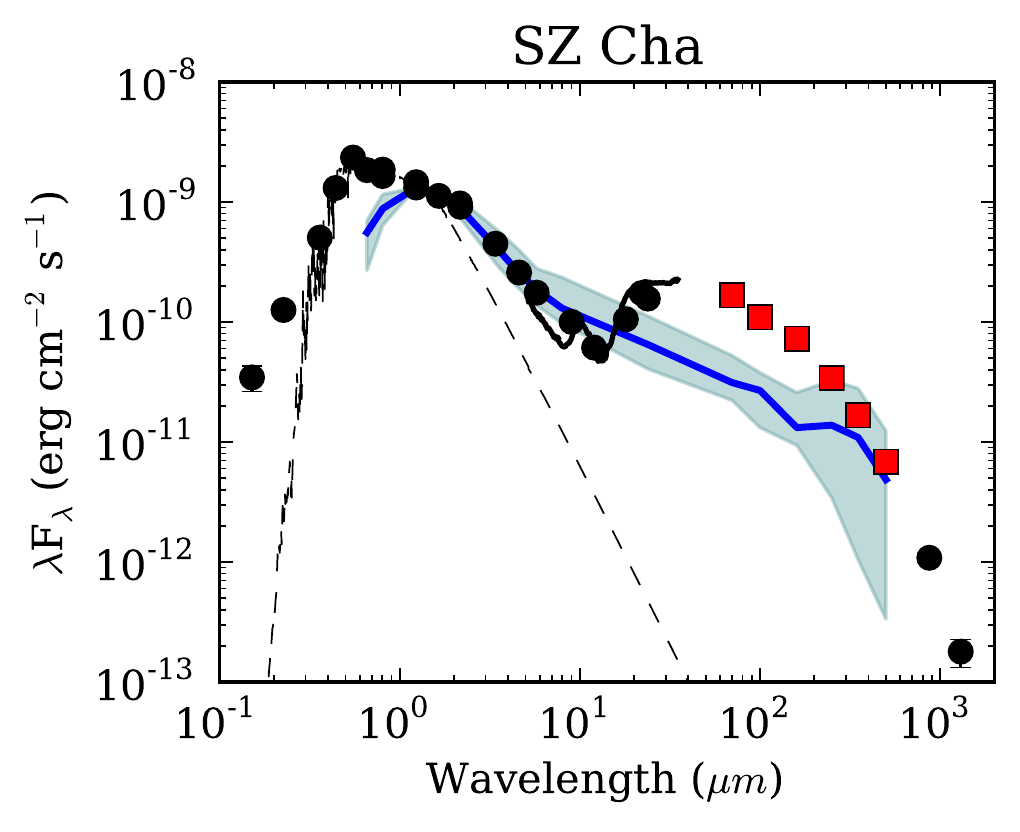}
  \includegraphics[width=0.32\hsize]{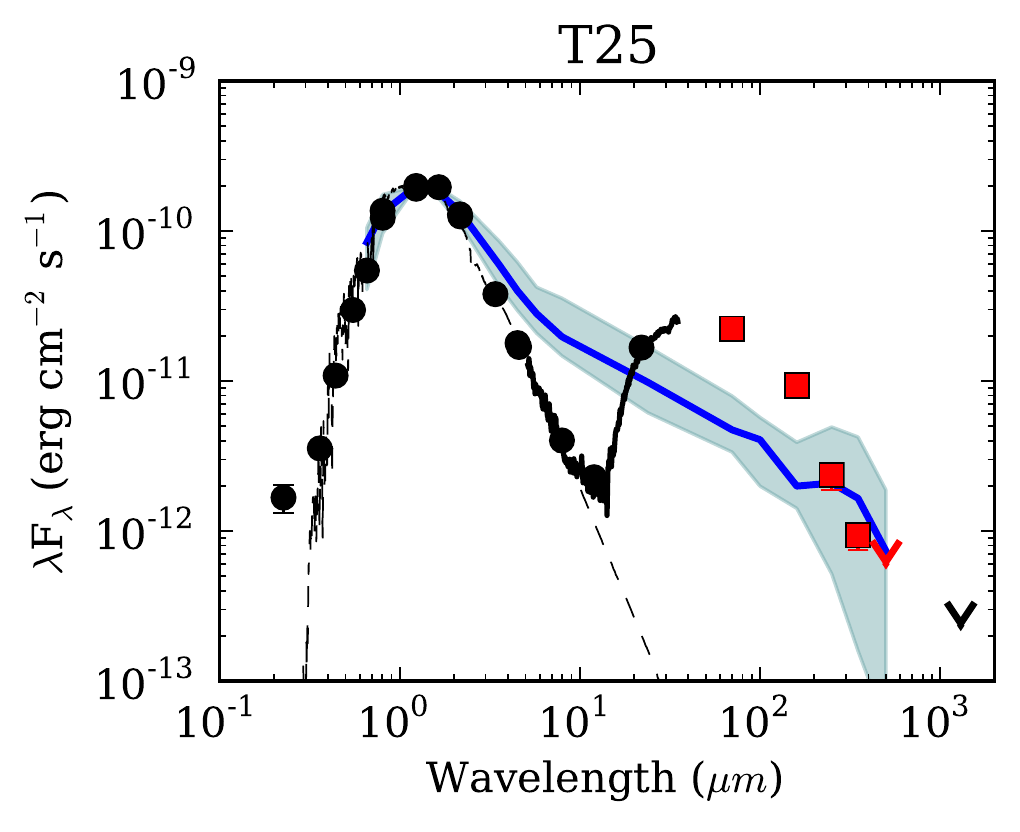}
  \includegraphics[width=0.32\hsize]{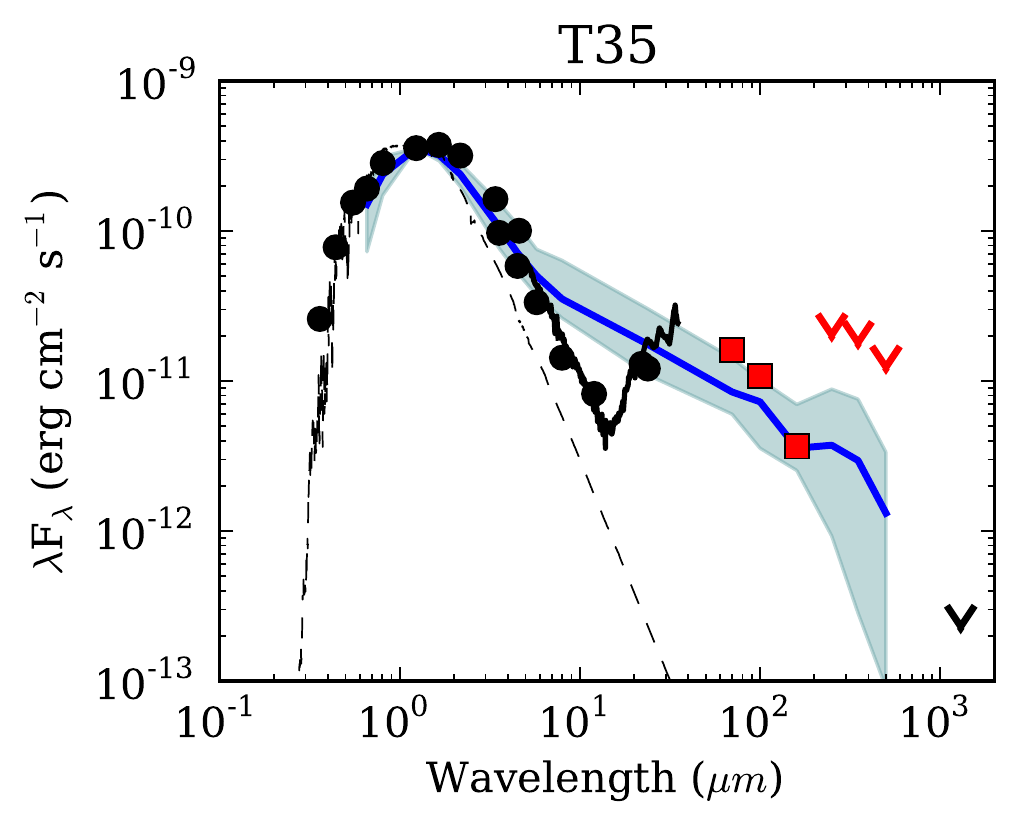}
  \includegraphics[width=0.32\hsize]{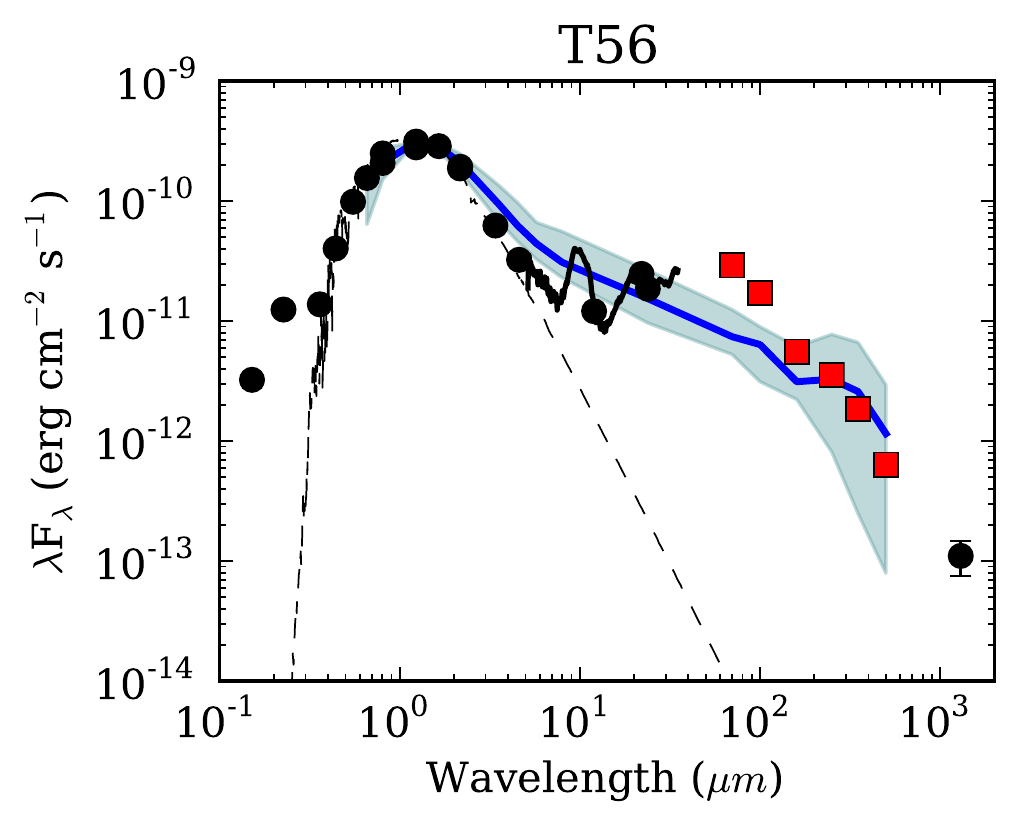}
  \includegraphics[width=0.32\hsize]{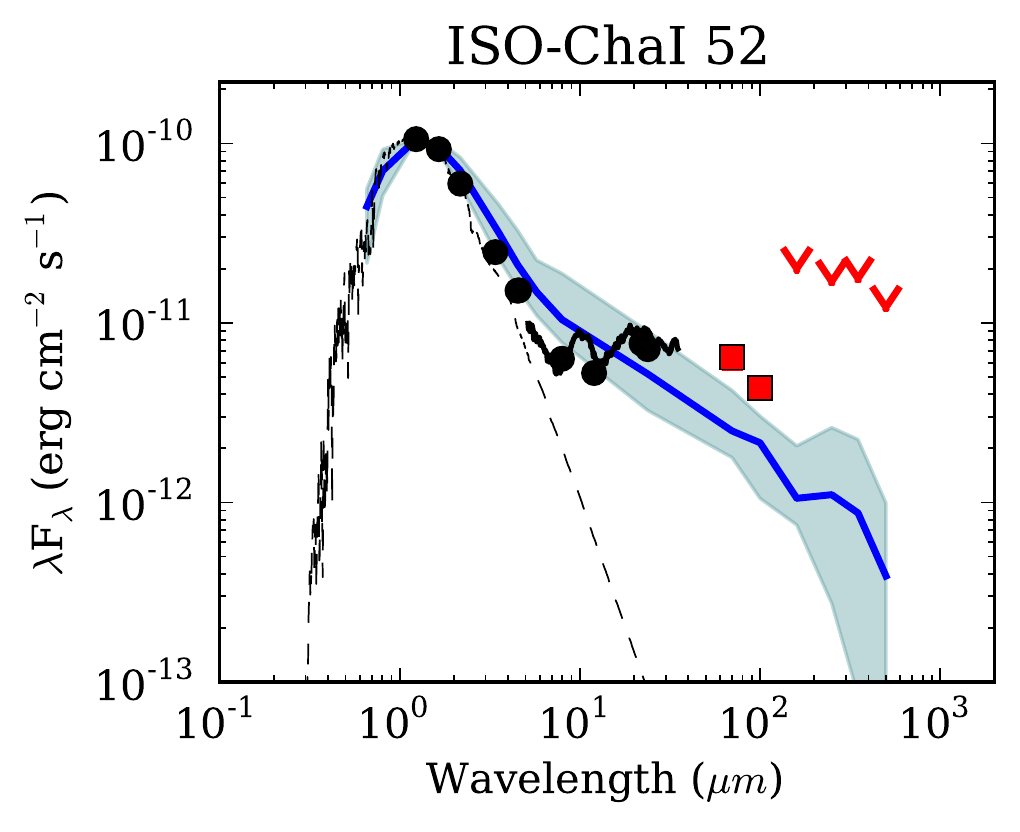}
  \caption{SEDs of the detected transitional disk candidates, confirmed by our classification criterion and updated with the fluxes from \textit{Herschel}. Black dots are the dereddened observed values from the literature, downward black arrows are flux upper limits from the literature. \textit{Herschel} data are represented in red (squares for detections, downward arrows for upper limits). Uncertainties are within symbol sizes. The IRS spectra from \citet{Manoj2011} (black solid lines) and the photospheres \citep[dashed black lines from the NextGen models from][]{Allard2012} are also plotted. The median Class II SED (blue solid line) and the first and fourth quartiles (blue area) are shown (see also Sect.~\ref{outerdisks} and Table~\ref{table:median}).}
\label{fig:sedsdetections}
\end{figure*}

\begin{figure*}
  \centering
  \includegraphics[width=0.32\hsize]{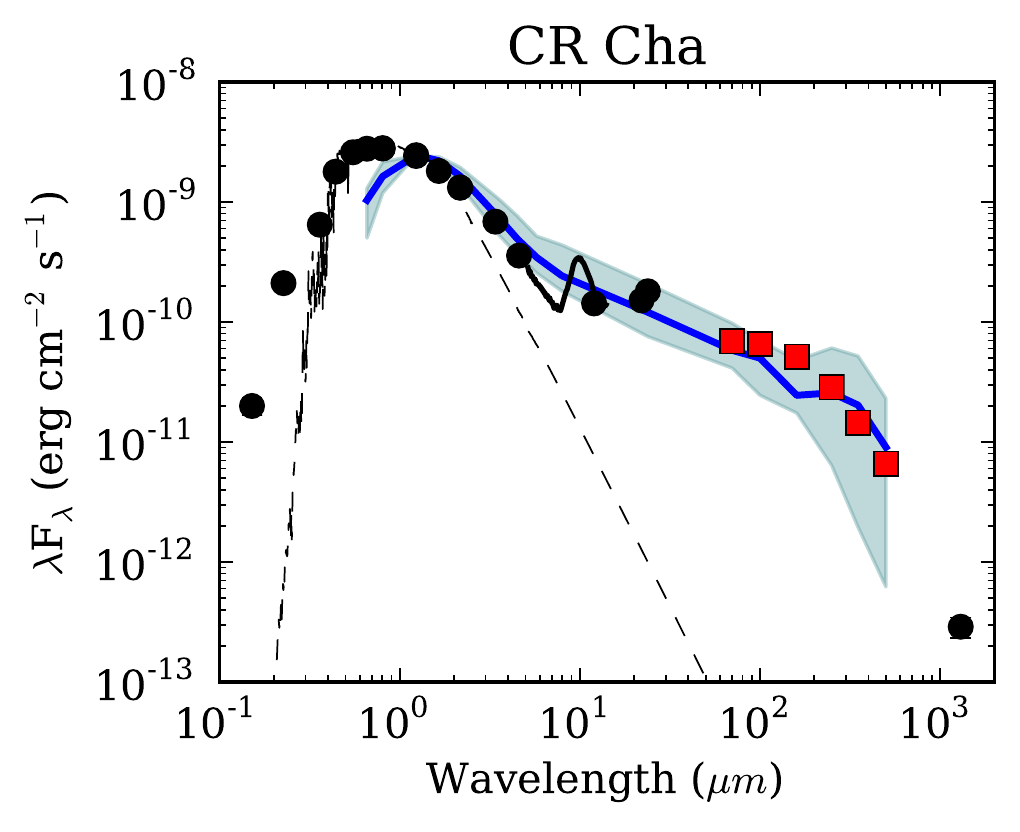}
  \includegraphics[width=0.32\hsize]{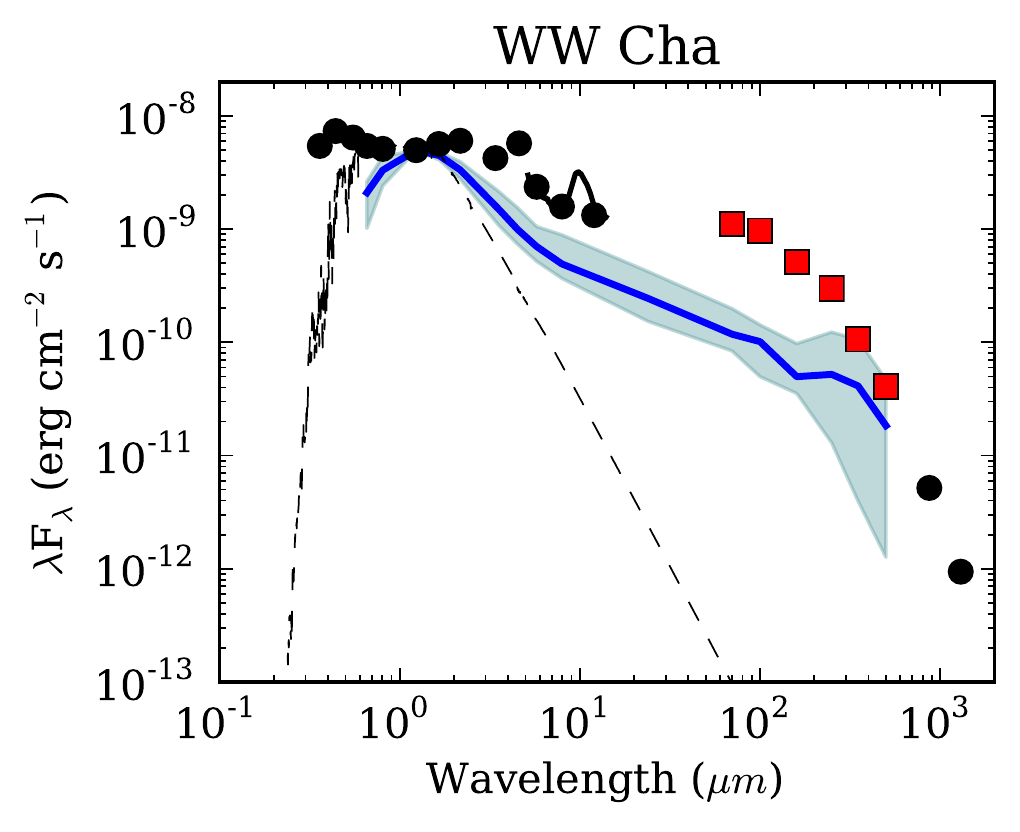}
  \includegraphics[width=0.32\hsize]{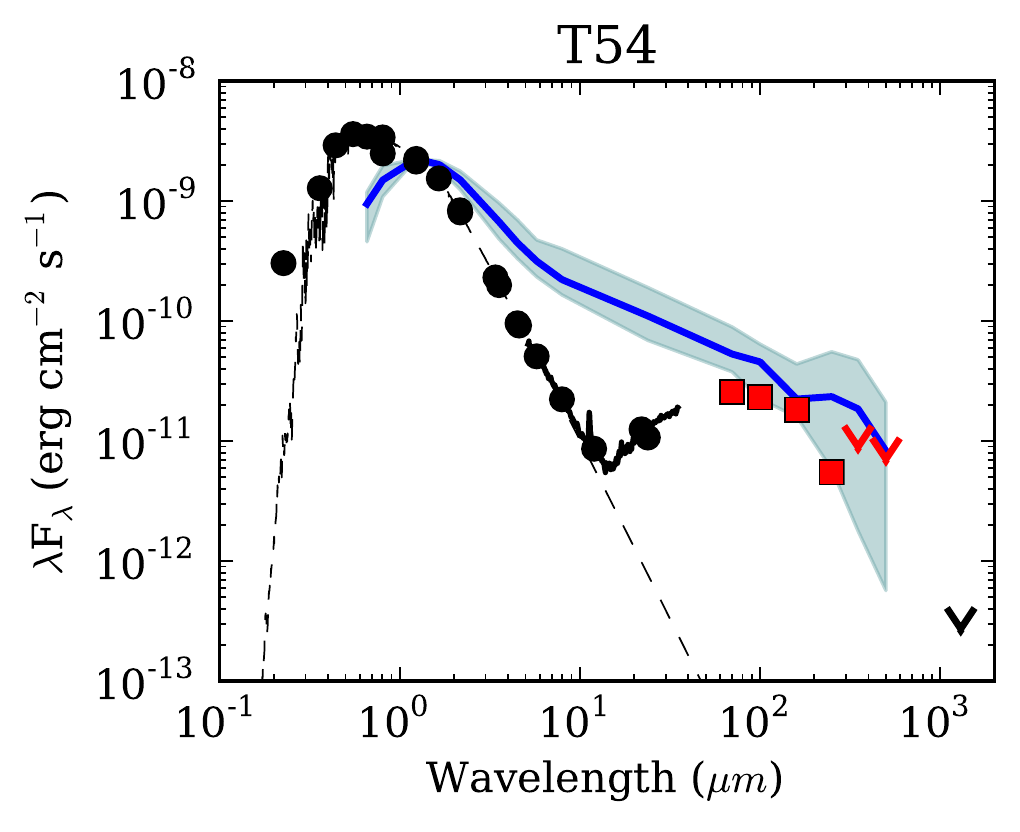}
  \caption{SEDs of transitional disk that do not fulfill our classification criteria. CR Cha and WW Cha display clear Class II SED. T54 appears extended in the \textit{Herschel} images. Symbols are as in Fig.~\ref{fig:sedsdetections}.}
  \label{fig:sedsmiss}
  \end{figure*}

\begin{figure*}
  \centering
  \includegraphics[width=0.32\hsize]{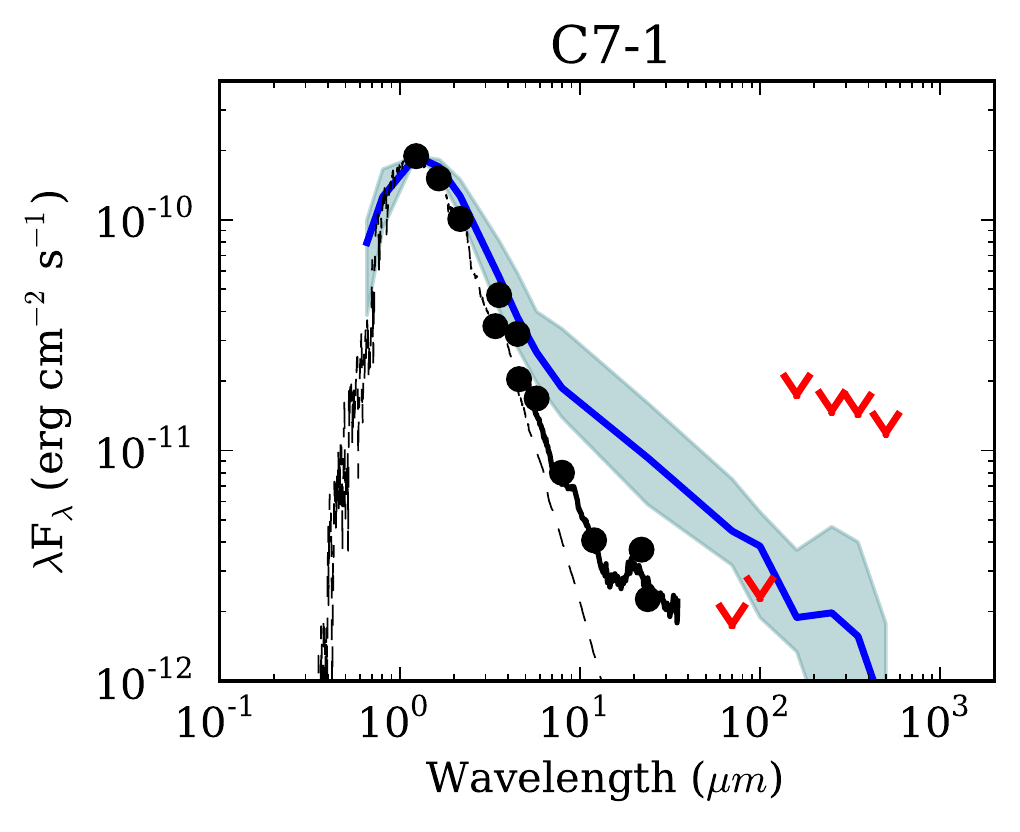}
  \includegraphics[width=0.32\hsize]{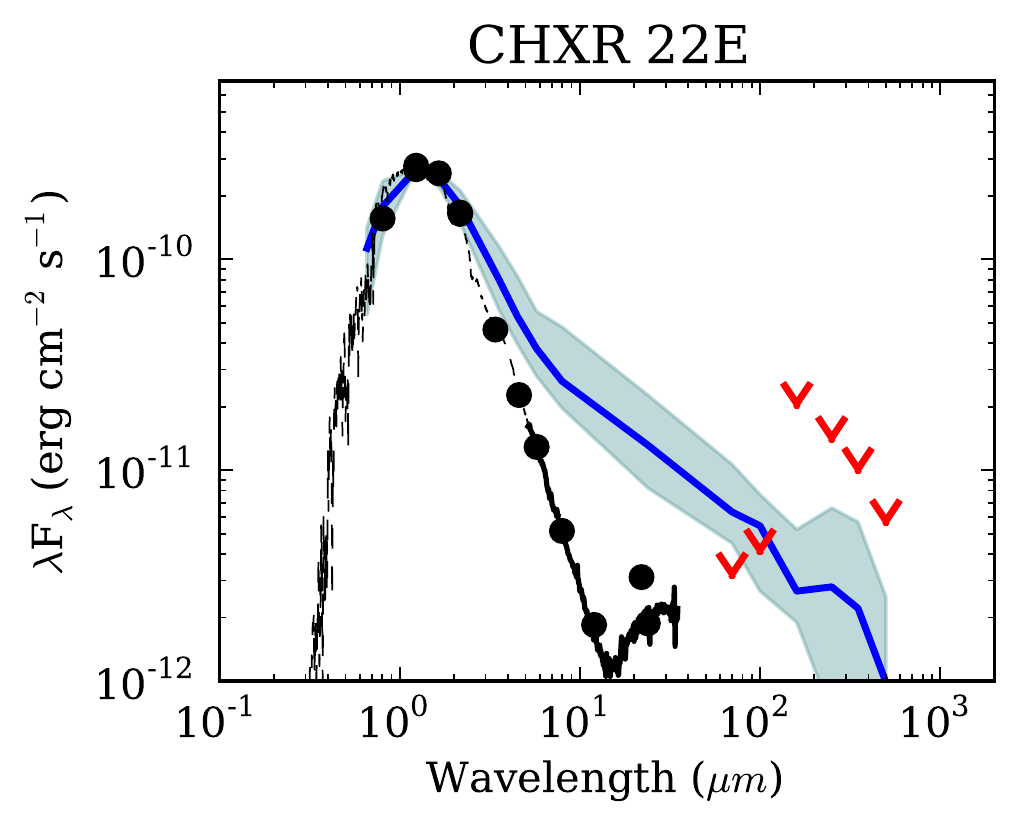}
  \includegraphics[width=0.32\hsize]{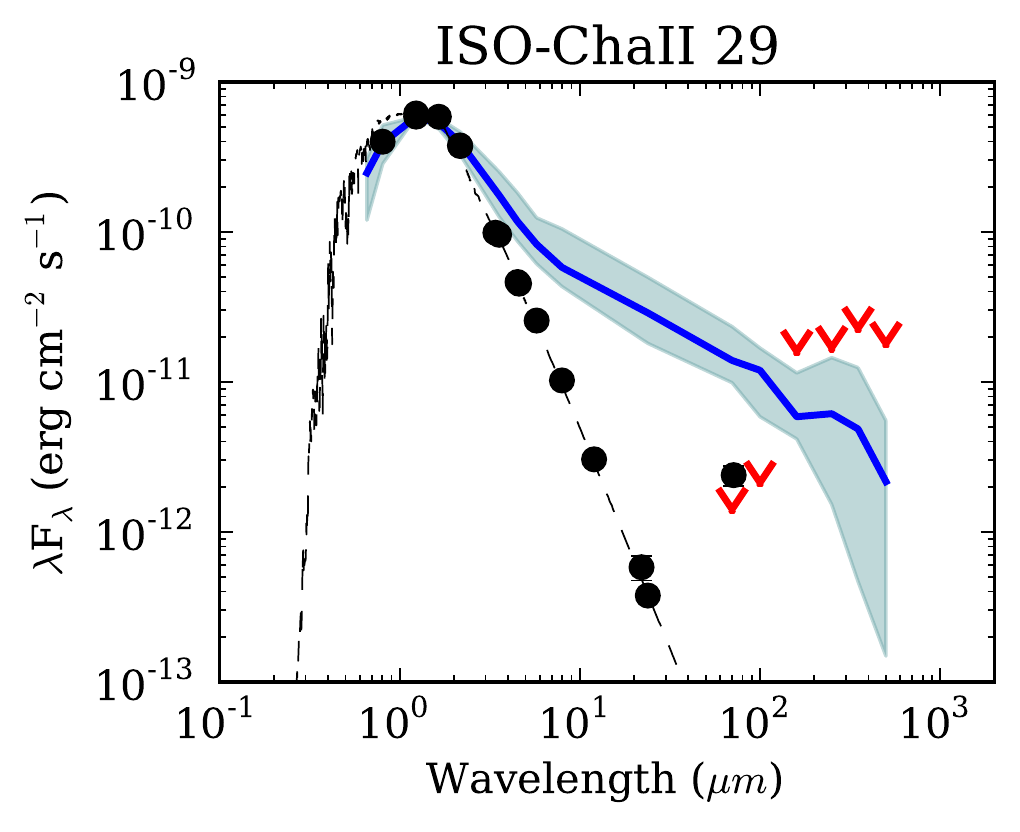}
  \caption{SED of the transitional disk candidates not detected by \textit{Herschel} for which only flux upper limits could be estimated. Symbols are the same as in Fig.~\ref{fig:sedsdetections}. The upper limits at 70\,$\mu$m for C7-1 and ISO-ChaII 29 allow us to classify them as non-transitional with our selection criteria.}
\label{fig:sedsnondetections}
\end{figure*}

\section{Transitional disks in the sample}\label{outerdisks}

\begin{table*}
  \caption{Normalized flux densities of the median SED, upper SED (fourth quartile) and lower SED (first quartile) of the Class II objects in Cha I and Cha II, after extinction correction. Transitional disks are not included. The number of stars detected in each band is also indicated. For comparison, we also include the median value for the six transitional disks detected in this study, although we stress its very low number statistics.}\label{table:median}\begin{center}
    \begin{tabular}{l c c c c c}
      \hline
      \hline\rule{0mm}{3mm}$\lambda$\,($\mu$m) & Median $_{\rm transitional}$ & Median & Upper & Lower & Detections\\
      \multicolumn{6}{c}{(F$_{\lambda}$ arbitrary units)}
      \\\hline
     \textit{R}  &  1.01 & 0.79  &  1.00  &  0.38 & 32\\ 
      \textit{I}  &  1.22 & 1.02  &  1.34  &  0.74 & 32\\ 
      \textit{J} &  1.00 & 1.00  &  1.00  &  1.00 & 107\\ 
      \textit{H}  &  0.67 & 0.67  &  0.72  &  0.61 & 107\\ 
      \textit{Ks} &  0.35 & 0.38  &  0.45  &  0.32 & 107\\ 
      IRAC 3.6  &  7.1e-2 & 0.10  &  0.15 &  7.5e-2 & 78\\ 
      IRAC 4.5  &  3.2e-2 & 5.5e-2  &  8.5e-2 &  4.0e-2 & 70\\ 
      IRAC 5.8  &  2.0e-2 & 3.0e-2  &  4.5e-2 &  2.2e-2 & 86\\ 
      IRAC 8.0  &  4.7e-3 & 1.5e-2  &  2.7e-2 &  1.1e-3 & 78\\ 
      MIPS 24  &  3.5e-3 & 2.5e-3  &  4.4e-3  &  1.6e-4 & 95\\ 
      PACS 70  &  1.8e-3 & 4.2e-4  &  7.0e-4  &  3.0e-4 & 23\\ 
      PACS 100  &  6.8e-4 & 2.5e-4  &  3.5e-4  &  1.2e-4 & 41\\ 
      PACS 160  &  3.5e-4 & 7.7e-5  &  1.5e-4  &  5.5e-5 & 19\\ 
      SPIRE 250  &  5.7e-5 & 5.1e-5  &  1.2e-4  &  1.3e-5 & 9\\ 
      SPIRE 350  &  1.9e-5 & 2.9e-5  &  7.5e-5  &  2.8e-6 & 9\\ 
      SPIRE 500  &  5.2e-6 & 9.1e-6  &  2.3e-5  &  6.3e-7 & 7
      \\\hline 
    \end{tabular}
  \end{center}
\end{table*}

We searched for additional photometric values in the literature for each of the transitional objects in our sample. \citet{Gauvin1992} reported optical photometry for all sources in Cha I except for CHXR 22E, ISO-ChaI 52, and C7-1. We have queried the VizieR catalog service and retrieved additional data for these targets from GALEX \citep{Galex}, 2MASS \citep{2MASS}, DENIS \citep{Denis}, WISE \citep{WISE}, and AKARI \citep{AKARI}. To avoid possible mismatching, we chose a search radius of 1\,\arcsec around the 2MASS coordinates. We rejected DENIS photometry for T25, T35, and ISO-ChaI 52, since it clearly disagreed with the rest of photometric data. (Sub)millimeter data at 870\,$\mu$m and 1.3\,mm were also included from \citet{Belloche2011} and \citet{Henning1993}, respectively. \textit{Spitzer} photometric measurements were included from \citet{Damjanov2007}, \citet{Luhman2008a}, \citet{Luhman2008b}, and \citet{Alcala2008}. We also retrieved the \textit{Spitzer} IRS spectra from the CASSIS database \citep{CASSIS}. The resulting SEDs for all sources are shown in Figs.~\ref{fig:sedsdetections}, \ref{fig:sedsmiss}, and~\ref{fig:sedsnondetections}. Thumbnails of the transitional disks as seen in the \textit{Herschel} 70\,$\mu$m images can be found in Fig.~\ref{fig:cha}. We note here that the cross-shaped PSF at 70\,$\mu$m is produced by the parallel mode observations, and does not represent resolved objects.

We compared the \textit{Herschel} fluxes of the sample of transitional disks with the Class II sources in the same region. For this purpose, we first inspected the SEDs of all Class II sources (after removing the transitional disks sample). We identified and removed one object (J11111083-7641574) previously classified as an edge-on disk \citep{Luhman2008b,Robberto2012}. It could not be identified in the slope-slope diagram since it is only detected at 100\,$\mu$m. The object is present in the \textit{Herschel} images at 70\,$\mu$m, but was not detected by \textit{Sussextractor} with the selected threshold. We built the median SED of all Class II objects, extinction corrected and normalized to the \textit{J}-band. We also computed SED of the first ($<$\,25\,\%) and fourth ($>$\,75\,\%) quartiles. Given the low detection numbers for the SPIRE bands, we used the lowest and highest values instead of the quartiles at those wavelengths. We included photometry from the \textit{R}, \textit{I}, 2MASS, \textit{Spitzer}, and \textit{Herschel} bands. We considered all objects detected in each band (regardless of whether they were undetected in the other bands). The obtained values are given in Table~\ref{table:median}. We also found the median SED not to vary significantly when only K or M type stars were considered.

The comparison between the SEDs of transitional disks and the median SED of Class II objects in Cha I and II shows a systematic difference in the 70-160\,$\mu$m range. The six transitional disks found with the selection criterion used in this study display a clear excess over the obtained Class II median SED, and all of them are over the fourth quartile level (uncertainties for T35 are consistent with a flux value below this level).

Even though this median SED was built with a relatively small statistical sample, this result clearlyshows that transitional disks are brighter at 70-160\,$\mu$m than typical Class II sources in these regions.

Similar phenomena were already tentatively described by \citet{Winston2012} in a preliminary study of the YSO population of Cha I and by \citet{Cieza2011TCha} for T Cha, but here we provide consistent results derived from a much larger sample of transitional disks. This excess was not previously found by large programs using the \textit{Spitzer} Space Telescope, such as \textit{cores 2 disks} \citep{Evans2003,Evans2009}. This can probably be explained by the lower sensitivity of \textit{Spitzer} at 70\,$\mu$m. 

A bias toward the brightest objects could affect these results in two different ways: we might miss the faintest transitional disks and the faintest Class II sources. In the first case, the \textit{Herschel} data are enough to classify eleven out of the twelve previously known transitional objects in the sample (less than 10\,\% objects missed). This suggests that the proposed method does not suffer from a strong bias effect. The existence of an unknown population of transitional disks not identified with \textit{Spitzer} cannot be ruled out \citep[although this possibility is unlikely, see][]{Merin2010}. However, this would not alter the systematic difference found at the 70-160\,$\mu$m range between the detected previously known transitional disks and Class II objects in these regions. On the other hand, the second scenario (e.g. missing faint Class II sources) would imply lower values for the Class II median SED in the \textit{Herschel} range, producing an even stronger difference between transitional and Class II disks. As a result, the Class II median SED computed here should be considered as an upper limit.

If the excess at PACS bands is a common feature of transitional objects, two explanations can be proposed to explain it: (1) transitional disks are not the result of the evolution of Class II sources, but a parallel evolutionary path, or (2) the discrepancy between transitional disks and Class II objects is produced during the transitional phase (maybe even by the same mechanism that causes the transitional disk evolution). In this case, the piling-up of mass at some point of the outer disk could produce the steep slope and the excess found at 70\,$\mu$m \citep{Williams2011}. With the available data it is not possible to favor any of these scenarios, so we leave this question open to future studies.

A larger and statistically more significant sample of transitional disks and modeling are required to confirm whether the difference found at the PACS bands applies to the whole transitional disk sample, to identify the real cause (or causes) of the excess, and to understand whether transitional disks are indeed a later stage of Class II objects or follow a different evolutionary path.

\section{Conclusions}\label{conclusions}

We presented a new method for identifying transitional disks based on the slope between the WISE 12\,$\mu$m and PACS 70\,$\mu$m bands. We have applied this method to the whole sample of known Class II objects in the Cha I and II star-forming regions. We were able to separate known transitional sources from Class II objects, and reclassified five objects as possibly non-transitional. This method could produce false positives due to the presence of edge-on disks, and Hertzsprung–Russell diagrams should be used to reject underluminous sources. As a result, we found an observed contamination level of $\sim$\,45\% among previously identified transitional disks in these regions. The size of our sample is relatively small, and these results cannot be applied to the whole transitional disk sample. However, a revision of the transitional disk population in other star-forming regions is warranted to determine the real contamination level and to account for its effects. 

We built the median SED of Class II sources in the regions, and found significantly higher PACS fluxes in the transitional disks compared to it. This excess could be produced during the transitional phase, or be explained in terms of a different evolutionary path for transitional disks and Class II sources.

Future studies of other star-forming regions observed by the \textit{Herschel} Space Observatory will clarify the contamination level of the sample of known transitional objects, and will provide stronger evidence for a systematic excess at PACS wavelengths in transitional disks with respect to Class II sources.

\begin{acknowledgements}

We thank the referee for his/her constructive comments. This work has been possible thanks to the ESAC Science Operations Division research funds, support from the ESAC Space Science Faculty and of the Herschel Science Centre. NLJC acknowledges support from the Belgian Federal Science Policy Office via the ESA's PRODEX Program. PACS has been developed by a consortium of institutes led by MPE (Germany) and including UVIE (Austria); KUL, CSL, IMEC (Belgium); CEA, OAMP (France); MPIA (Germany); IFSI, OAP/AOT, OAA/CAISMI, LENS, SISSA (Italy); IAC (Spain). This development has been supported by the funding agencies BMVIT (Austria), ESA-PRODEX (Belgium), CEA/CNES (France), DLR (Germany), ASI (Italy), and CICT/MCT (Spain). SPIRE has been developed by a consortium of institutes led by Cardiff Univ. (UK) and including Univ. Lethbridge (Canada); NAOC (China); CEA, LAM (France); IFSI, Univ. Padua (Italy); IAC (Spain); Stockholm Observatory (Sweden); Imperial College London, RAL, UCL-MSSL, UKATC, Univ. Sussex (UK); and Caltech, JPL, NHSC, Univ. Colorado (USA). This development has been supported by national funding agencies: CSA (Canada); NAOC (China); CEA, CNES, CNRS (France); ASI (Italy); MCINN (Spain); SNSB (Sweden); STFC (UK); and NASA (USA).This study also makes use of the data products from the Two Micron All Sky Survey (2MASS), a joint project of the University of Massachusetts and IPAC/Caltech, funded by NASA and the National Science Foundation; data products from the Wide-field Infrared Survey Explorer (WISE), a joint project of the University of California, Los Angeles, and the Jet Propulsion Laboratory (JPL)/California Institute of Technology (Caltech); data produts from DENIS, a project partly funded by the SCIENCE and the HCM plans of the European Commission under grants CT920791 and CT940627; the NASA Infrared Processing and Analysis Center (IPAC) Science Archive; the SIMBAD database; and the Vizier service, operated at the Centre de Données astronomiques de Strasbourg, France.

\end{acknowledgements}

\bibliographystyle{aa}
\bibliography{mybiblio}

\begin{appendix}
\section{Literature review of the individual transitional objects detected with \textit{Herschel}}\label{sec:appendix}

\subsection{CS Cha}

CS Cha was first studied by \citet{Gauvin1992}, who found evidence that it harbors a disk with inner holes. It is known to be a spectroscopic binary system, as confirmed by \citet{Guenther2007} (period $\ge$ 7\,years, minimum mass of the companion $\sim$0.1\,M$_{\odot}$), although previously \citet{Takami2003} suggested this option based on the large gap found in its disk. This previously unknown feature is probably the reason for the spectral type inconsistency found in the literature \citep{Henize1973,Appen1977,Ryd1980,Appen1983,Luhman2004}. In this study we used the K6 spectral type found by \citet{Luhman2007}. The binary nature makes the disk around CS Cha into a circumbinary disk. The disk has been modeled intensively, initially excluding the effect of the binary system \citep{Espaillat2007b,Espaillat2011} and evidence of an inner hole of $\sim$\,40\,AU was found. \citet{Espaillat2011} also pointed out the need for a different mass distribution in CS Cha compared to that of disks around single stars. A more recent analysis by \citet{Nagel2012} also accounted for the binary effect. To reproduce the variations found at the IR wavelengths, the model includes the emission from the inner disk structure generated by the double system, with a ring and streams of material falling from the ring to the circumstellar disks around the individual stars. Another 2MASS source is found at 5\arcsec away. It is 6 magnitudes weaker than CS Cha in the 2MASS J band and undetected in the rest of the 2MASS bands. Contamination from this object is therefore very unlikely.

CS Cha is located in front of a bright background. Therefore, the SPIRE fluxes are very likely underestimated because the background emission was probably overestimated during the photometry extraction.

\subsection{SZ Cha}

This source was cataloged as a K0 star by \citet{Ryd1980} and was first identified as a disk with a possible inner gap by \citet{Gauvin1992}. \citet{Luhman2007} reviewed its properties, and it was lately confirmed by \citet{Kim2009} as a transitional disk. It has sometimes been referred to as a pretransitional disk, given the small excess found at 3-10\,$\mu$m. The first modeling results by \citet{Kim2009} suggested an inner disk radius of $\sim$\,30\,AU. \citet{Espaillat2011} modeled this object in detail, noting flux variations from IRS spectra at different epochs on periods shorter than three years. These variations are attributed to changes in the height of the optically thick disk wall (from 0.006 to 0.08\,AU), and they do not modify the 10\,$\mu$m silicate emission feature. SZ is known to be a wide binary \citep{Vogt2012}. A companion is found at $\sim$\,5\arcsec (projected distance of 845\,AU), which could be causing truncation of the outer disk. The contribution of this source to the total measured fluxes in this study is likely to be negligible, since it is 4 magnitudes weaker than SZ Cha in the 2MASS J band. However, the possibility of an increase in its FIR measurements cannot be excluded.

\subsection{T25}

T25 was identified as an M3 star by \citet{Luhman2008a} and was found to be a transitional disk by \citet{Kim2009}. The lack of IR excess at wavelengths $<$\,8\,$\mu$m indicates that the inner regions of the disk are well depleted of small dust particles. The modeling by \citet{Kim2009} yields an inner radius of 8\,AU for the disk. It is the only detected transitional object, together with T35, lacking the silicates feature at 10\,$\mu$m, another indication of an efficient depletion of small particles in the inner disk region. T25 has no known stellar companions \citep{Nguyen2012}.

\subsection{T35}

\citet{Gauvin1992} classified this source as an M0 star, and it was later identified as a possible a pretransitional disk by \citet{Kim2009} because it displays weak excess at short IR wavelengths. In this case, the inner disk radius is located at 15\,AU \citep{Espaillat2011}. As in T25, there is no sign of silicate emission. The excess at 70\,$\mu$m is lower than in other cases, but does not resemble the typical Class II SED. It has no confirmed known stellar or substellar companions \citep{Nguyen2012}. However, recent sparse aperture masking observations of this source by \citet{Cieza2013} showed and asymmetry in its K-band flux. On the basis of modeling, these authors found the inner disk radius to be $\sim$8.3\,AU. They were also unable to distinguish between the close-companion scenario and the asymmetry being produced by the starlight scattered off the disk itself.

\subsection{T56}

This source was found to be an M0.5 start in \citet{Gauvin1992}. \citet{Kim2009} identified it as a transitional disk with a inner disk radius of 18\,AU. As in the other transitional disks in this study, its excess is higher than the expected Class II flux at the PACS bands. It has no known bound companions \citep{Nguyen2012}.

\subsection{ISO-ChaI 52}

ISO-ChaI 52 is an M4 star \citep{Luhman2004}. \citet{Espaillat2011} proposed it as a transitional disk, finding the source to be an extreme case among their sample: based on variations of its \textit{Spitzer} spectrum, models require the inner wall height to increase by about 400\,\% (from 0.0006 to 0.003\,AU). We also found it to be an outlier in the sense that it has the flattest SED between 12 and 70\,$\mu$m. No bound companions are known for ISO-ChaI 52 \citep{Nguyen2012}.

\subsection{CR Cha}

CR Cha is an M0.5 star \citep{Gauvin1992}. \citet{Furlan2009} found it to be an outlier in their sample when comparing the equivalent width of the silicates emission and the spectral slope between 13 and 31\,$\mu$m: it was beyond the parameter space considered in their study. The explanation given in \citet{Furlan2009} is that this source could be a pretransitional disk. For this reason, \citet{Espaillat2011} included it in their sample of transitional disks. In this study, we found this object to be compatible with a Class II object. It is also located among other Class II objects using the proposed classification method (Fig.\ref{fig:slopes}). Therefore, although we cannot completely rule out the possibility that this object is in a pretransitional disk phase given its strong silicates emission, it would be in a very early stage of the transitional phase.

\subsection{WW Cha}

This source was first classified as a K5 object by \citet{Gauvin1992}. It was included in the analysis of \citet{Espaillat2011} for the same reason as CR Cha, and modeled as a pretransitional disk. Comparison with the median SED of the Class II sources shows that WW Cha is well above the median. The SED of WW Cha resembles a typical Class II object, probably still embedded in the core, as suggested by its high extinction ($A_v\sim$ 4.8 mag) and its position in the \textit{Herschel} maps. The dusty environment  in which it is located could significantly pollute the photometry and, hence, our conclusions about this object.

\subsection{T54}

T54 is known to be a misclassified transitional disk \citep{Matra2012}, and therefore we excluded it from our analysis. The \textit{Herschel} images show contamination from close-by extended emission, which affected our photometry and, hence, our conclusions. The non-transitional nature of this object is also supported by the fact that it would be the only transitional disk in our sample with no excess emission at 70\,$\mu$m with respect to the median SED Class II disks.

\end{appendix}

\end{document}